\documentclass[aps,prl,showpacs,superscriptaddress,twocolumn,final,numerical,10pt]{revtex4-1}
\usepackage{epsfig}
\usepackage{graphicx}
\usepackage{dcolumn}
\usepackage{bm}
\usepackage{amsthm}
\usepackage{amsfonts}
\usepackage{amscd}
\usepackage{amsmath}
\usepackage{enumerate}
\usepackage{bbm}

\newcommand{\ket}[1]{\mbox{$ | #1 \rangle $}}
\newcommand{\bra}[1]{\mbox{$ \langle #1 | $}}
\newcommand{\braket}[2]{\ensuremath{\langle #1|#2\rangle}}

\DeclareMathOperator{\tr}{tr}
\newcommand{\ie}{\textit{i.e.}}
\newcommand{\eg}{\textit{e.g.}}
\newcommand{\st}{\textrm{s.t.}}

\begin{document}

\title{Security of distributed-phase-reference quantum key distribution}

\author{Tobias Moroder}
\affiliation{Naturwissenschaftlich-Technische Fakult\"at, Universit\"at Siegen, Walter-Flex-Str.~3, D-57068 Siegen, Germany}
\author{Marcos Curty}
\affiliation{EI Telecomunicaci\'on, Dept. of Signal Theory and Communications, University of Vigo, E-36310 Vigo, Spain}
\author{Charles Ci Wen Lim}
\affiliation{Group of Applied Physics, University of Geneva, CH-1211 Geneva, Switzerland}
\author{Le Phuc Thinh}
\affiliation{Centre for Quantum Technologies, National University of Singapore, 3 Science Drive 2, Singapore 117543, Singapore}
\author{Hugo Zbinden}
\affiliation{Group of Applied Physics, University of Geneva, CH-1211 Geneva, Switzerland}
\author{Nicolas Gisin}
\affiliation{Group of Applied Physics, University of Geneva, CH-1211 Geneva, Switzerland}

\date{\today}

\begin{abstract}
Distributed-phase-reference quantum key distribution stands out for its easy implementation with present day technology. Since many years, a full security proof of these schemes in a realistic setting has been elusive. For the first time, we solve this long standing problem and present a generic method to prove the security of such protocols against general attacks. To illustrate our result we provide lower bounds on the key generation rate of a variant of the coherent-one-way quantum key distribution protocol. In contrast to standard predictions, it appears to scale quadratically with the system transmittance.
\end{abstract}

\maketitle


\textit{Introduction.}---
Quantum key distribution (QKD) is on the verge to become a standard tool for secure communications~\cite{qkd}. In its original proposal QKD is based on the transmission of single photons. However, since true single photon sources are not available yet most experimental prototypes and all current commercial products of QKD use weak laser pulses. A main drawback of these systems is that some signals contain more than one photon prepared in the same quantum state. This fact severely limits the distances that can be achieved by these techniques due to the photon number splitting attack~\cite{pns}.

To enhance the performance of practical QKD systems, several approaches have been proposed. One solution is to send a strong reference pulse together with the quantum signals~\cite{strong}. A second approach is based on the decoy state method, where the transmitter sends states with different intensities~\cite{decoy}. Both schemes provide a secret key generation rate that scales linearly with the system transmittance~\cite{strong,decoy}. A third alternative is to use distributed-phase-reference (DPR) QKD protocols. They differ from standard QKD schemes in that the receiver now performs joint measurements onto subsequent signals, often given in the form of coherence measurements~\cite{dps,cow}. This approach includes the differential-phase-shift \cite{dps} and the coherent-one-way (COW) \cite{cow} protocols. In the former, the sender prepares coherent states of equal intensity but modulates their phases; in the COW protocol all pulses share a common phase but their intensities vary. A complete security proof of DPR-QKD in a realistic setting has been missing since many years. Security has only been proven so far against restricted types of attacks~\cite{dps_individualattack, dps_sequential,upper_bound_cow}, or assuming the use  of ideal single photon sources~\cite{kiyo}.

In this Letter we present a generic method to prove security of practical DPR-QKD against general attacks. This solves a long standing open question in the field of quantum communications \cite{qkd}. We illustrate our result by providing non-trivial lower bounds for a variant of the original COW protocol~\cite{private}, which maintains all the practical advantages. Our analysis suggests that practical DPR-QKD might not be as robust against imperfections as initially foreseen, \ie, its key rate appears to scale quadratically with the system transmittance.

\textit{Security discussion.}---
The challenge in DPR-QKD is to prove security against general, also termed coherent attacks. Usually such attacks are known to be of no advantage to the eavesdropper (Eve) in comparison to collective attacks by virtue of the de Finetti theorem~\cite{de_finetti}. This theorem applies, for instance, when the underlying quantum state shared by the legitimate users (Alice and Bob) is permutationally invariant. In standard QKD this is typically ensured by performing simultaneous random permutations on the classical measurement results. DPR-QKD defines however a fixed ordering of the signals by its coherence measurement and, therefore, it is not possible to permute the classical outcomes without destroying vital information~\cite{note_friday}. However, such a predicament can be circumvented by grouping the entire signal stream into blocks. More specifically, consider that Alice and Bob group their signals into subsequent blocks of size $m$, where the length $m$ is optimized for the expected behaviour. 
When permuting these blocks one preserves the coherence information within them while the information between the blocks is destroyed. Still this is enough to apply the de Finetti argument on the level of blocks. As a result, the state shared by Alice and Bob after distributing a large number $mN$ of signals satisfies $\rho_{\rm AB}^{mN}\approx \rho_{\rm AB}^{m\;\otimes N}$ and security against collective attacks on these signal blocks implies security against coherent attacks in the original setting.

Suppose that the state shared by Alice, Bob and Eve after transmitting an $m$ block signal is $\rho_{\rm ABE}^m$. Let us first consider the effect of public announcements by Alice and Bob based on their classical measurement results. This announcement, labelled as $v$, allows both parties to distinguish between conclusive events that contribute to the sifted key and inconclusive ones that are discarded. On the level of quantum states this is described by suitable maps $\Lambda^{\rm A}_v \otimes \Lambda^{\rm B}_v$. Given an announcement $v$, that happens with probability $p(v)$, the three parties share the state $\sigma_{{\rm \bar A\bar BE},v}^{m}$ determined by $\Lambda^{\rm A}_v \otimes \Lambda^{\rm B}_v(\rho_{\rm ABE}^m) = p(v) \sigma_{{\rm \bar A \bar B E},v}^{m}$. 

For each announcement $v$ one can use a one-way classical post-processing key rate formula~\cite{lower_bounds}. If system ${\bar A}$ denotes a qubit  and Alice's raw key is obtained by projecting this system onto the orthogonal states $\ket{0}_{\rm \bar A}, \ket{1}_{\rm \bar A}$, then a lower bound on the secret key rate is given by $1-h_2(e_v)-h_2(\delta_v)$. Here $h_2$ represents the binary entropy, $e_v$ is the symmetrized bit error between the key measurements of Alice and Bob, and $\delta_v$ denotes the corresponding error, typically called phase error, when Alice performs a measurement in a mutually unbiased basis and Bob in his other different setting. This last parameter is used to upper bound Eve's knowledge on the sifted key generated by Alice. Note that $\delta_v$ does not need to be measured directly, it only needs to be estimated. When Bob's measurements are similar qubit measurements like the ones of Alice then the expression above represents the Shor-Preskill key rate formula~\cite{shor00a}. 

To consider that the output system ${\bar A}$ is a qubit implies that Alice can, at best, distill one secret bit per block. Nevertheless this restriction should not have a significant impact on the key rate in a long distance regime, since Bob observes, if any, most often only one single conclusive event per $m$ arriving signals due to the high losses in the channel (given that $m$ is not too big).

Instead of estimating separate phase errors $\delta_v$, it is often easier to combine all conclusive announcements $v \in \mathcal{V}_{\rm c}$ into an averaged version. Let $G=\sum_{v \in \mathcal{V}_{\rm c}}~p(v) \leq 1$ denote the total sifted key gain. Then, we have that the secret key rate per block can be bounded by
\begin{eqnarray}
\label{eq:bound_individual}
R_m &\geq& \inf_{\rho_{\rm ABE}^m}  \sum_{v \in \mathcal{V}_{\rm c}} p(v) \left[ 1-h_2( e_v) - h_2 (\delta_{v}) \right] \\
\nonumber
&\geq& \inf_{\rho_{\rm ABE}^m} G \left[ 1-h_2( \bar e_c) - h_2 (\bar \delta /G) \right] \\
\label{eq:bound_average}
&\geq& G \left[ 1-h_2 (\bar e_c) -h_2( \bar \delta^{\rm max}/G) \right].
\end{eqnarray}
Here one uses concavity of $h_2$ to lower bound $R_m$ by the averaged (conditional) error rates $\bar e_c=\sum_{v \in \mathcal{V}_{\rm c}} p(v) e_v /G$ and $\bar \delta= \sum_{v \in \mathcal{V}_{\rm c}} p(v) \delta_v$. The last step takes into account that $\bar e_c$ and $G$ are observed quantities and that the optimization is attained at the largest phase error $\bar \delta^{\rm max}$ compatible with the obtained data since $h_2$ increases in $[0,\frac{1}{2}]$.

\textit{Phase error estimation.}---
The main difficulty to compute Eq.~(\ref{eq:bound_average}) is to upper bound the average phase error $\bar \delta$. This parameter can be expressed as an expectation value on the original bipartite state $\rho_{\rm AB}^m=\tr_{\rm E}(\rho_{\rm ABE}^m)$ using adjoint maps
\begin{eqnarray}
\nonumber
\bar \delta & =& \sum_{v \in \mathcal{V}_{\rm c}} p(v) \tr(\sigma^{m}_{{\rm \bar A \bar B},v} F_{\delta_v})\! = \!\sum_{v \in \mathcal{V}_{\rm c}} \tr[\Lambda^{\rm A}_v \otimes \Lambda^{\rm B}_v (\rho_{\rm AB}^m) F_{\delta_v}\!] \\ 
\label{eq:phase_error_operator}
& =& \tr[\rho_{\rm AB}^m \sum_{v \in \mathcal{V}_{\rm c}} \Lambda^{{\rm A}\dag}_v \otimes \Lambda^{{\rm B}\dag}_v(F_{\delta_v})] = \tr(\rho_{\rm AB}^m F_{\bar \delta}).
\end{eqnarray}
Here $F_{\delta_v}$ denote the corresponding phase error operators on the state $\sigma^{m}_{{\rm \bar A \bar B},v}$. Partial knowledge of Alice and Bob about the state $\rho_{\rm AB}^m$ can be parsed as known expectation values $k_i=\tr(\rho_{\rm AB}^m K_i)$ for certain operators $K_i$. This means that the search for the maximum phase error $\bar \delta^{\rm max}$ can be cast into the form of a semidefinite program~\cite{sdp},
\begin{eqnarray}
\label{eq:phase_error_SDP}
\max && \:\tr(\rho_{\rm AB}^m F_{\bar \delta}) \\
\nonumber
\st && \: \rho_{\rm AB}^m \succeq 0,\: \tr(\rho_{\rm AB}^m K_i) = k_i\:\: \forall i .
\end{eqnarray}
Such special convex optimization problems can be solved efficiently using standard tools to obtain the exact optimum, even for large dimensions.

\textit{Available information and its description.}---
Let us be more precise about which expectation values $k_i$ are known in a prepare and measure scheme, where Alice sends potentially mixed states $\rho_i^m$ with a priori probability $p(i)$. This state preparation can be formulated in an entanglement based version as follows~\cite{mixed_states}: Alice first creates a source state $\ket{\Psi^m}_{\rm A_bA_sB} = \sum_i \sqrt{p(i)} \ket{i}_{\rm A_b}\ket{\rho_i^m}_{\rm A_sB}$, where $\ket{\rho_i^m}_{\rm A_sB}$ denote purifications of the signal states $\rho_i^m$ to a shield system ${\rm A}_{\rm s}$~\cite{shield}. Afterwards, she measures her bit system ${\rm A}_{\rm b}$ in the standard basis, thereby producing the correct signal state at site $\rm B$ which is sent to Bob. Eve transforms the overall source state to the final tripartite state $\rho_{\rm ABE}^m$ with ${\rm A}={\rm A}_{\rm b}{\rm A}_{\rm s}$. On the receiving side, Bob performs a measurement modelled by $B_k$. As a result, both Alice and Bob observe the expectation values of $\ket{i}_{\rm A_b}\bra{i}\otimes \mathbbm{1}_{\rm A_s} \otimes B_k$. Moreover, since Eve is restricted to interact only with Bob's system, the reduced density matrix $\rho_{\rm A}^m=\tr_{\rm BE}(\rho_{\rm ABE}^m)$ is fixed and directly given by the source state. This information can be added by including expectation values of $T_k \otimes \mathbbm{1}_{\rm B}$, where $T_k$ denotes a tomographic complete operator set on ${\rm A}$. Both sets of observables constitute the previously denoted set~$K_i$. 

The signal states and performed measurements in practical DPR-QKD are described by operators on an infinite dimensional Fock space of several modes. In order to apply the de Finetti argument~\cite{de_finetti}, and to numerically obtain an upper bound on the phase error using Eq.~(\ref{eq:phase_error_SDP}), it is necessary to formulate this problem in a manageable, finite dimensional form. Clearly, system $A_{\rm b}$ is finite. For Bob's measurements one can employ the squash model argument~\cite{squash}. Here the real measurement is notionally decomposed into a two step procedure by first applying a map that transforms any incoming signal to a finite dimensional output state on which a specified target measurement $B_k$ is performed afterwards. Since this map can be even given to Eve, its output state only lowers the key generation capabilities of Alice and Bob, and one readily works in finite dimensions. For our simulations we assume that Bob has at his disposal inefficient photon number resolving detectors with state independent dark counts. Also, we consider that only the single photon events within the whole block are finally considered as conclusive. In this case the map outputs either a single photon, measured with the perfect detection scheme, or an auxiliary state that triggers all inconclusive events. 

For the shield system $\rm A_s$ one uses only partial information of the reduced state. In the case of phase randomized signal blocks, an example that we consider later, a purification is given by storing the total photon number of the block in the shield system $\ket{n}_{\rm A_s}$. Using only tomography on the subspace spanned by all $n=1,\dots,n_{\rm cut}$, together with an ancilla state $\ket{N}_{\rm A_s}$ for all other cases, the shield system can effectively be described in finite dimensions.

\begin{figure}[!t]\center
\resizebox{8.6cm}{!}{\includegraphics{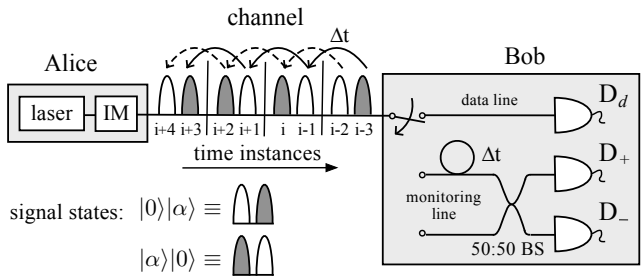}} \caption{
Schematic description of a COW protocol \cite{private} with an active measurement choice. Bob reads the raw key in detector ${\rm D}_d$. Moreover, he uses an optical switch to send some pairs of consecutive pulses to a monitoring line that examines the coherence between even and odd pulses sent by Alice.\label{COWfig}}
\end{figure}

\textit{Description of the protocol.}---
To illustrate our results we analyze the security of a variant of the COW protocol~\cite{private}. The basic setup is shown in Fig.~\ref{COWfig}. 
Alice uses a laser, followed by an intensity modulator (IM), to prepare a sequence of coherent states $\ket{0}\ket{\alpha}$ and $\ket{\alpha}\ket{0}$. On the receiving side, Bob employs an optical switch to distribute each pair of incoming pulses into the data or the monitoring line~\cite{comment_active}. The data line measures the arrival time of the pulses in detector ${\rm{D}}_d$ and creates the raw key. Whenever Bob sees a ``click" in this detector in say time instance~$i$, he decides at random whether to publicly announce a detection event in time instances $i$ and $i+2$ or $i$ and $i-2$. The first case is associated with a bit value ``0", while the second one corresponds to a bit value ``1". If the state sent by Alice in these time instances is $\ket{0}\ket{\alpha}$ ($\ket{\alpha}\ket{0}$) then she assigns to it a bit value ``0" (``1") and tells Bob to keep his result. Otherwise, the result is discarded and does not contribute to the sifted key. Let us illustrate this procedure with a simple example drawn in Fig.~\ref{COWfig}, and assume that Bob observes a click in ${\rm D}_d$ in time $i$. If he announces the pair $i$ and $i+2$, then this result is discarded. Note that in this case Alice sent $\ket{\alpha}\ket{\alpha}$ and hence she cannot infer in which time slot Bob saw a ``click". On the contrary, if Bob reports $i$ and $i-2$, then both parties assign to it a bit value ``1". The monitoring line checks for eavesdropping by measuring the coherence between subsequent even and odd pulses sent by Alice. This is done by interfering adjacent pairs of pulses in a $50:50$ beamsplitter and measuring the output states in detectors $\rm{D}_{+}$ and~$\rm{D}_{-}$. 

In the security analysis we assume that Alice and Bob discard coherence information between consecutive signal blocks. Moreover we consider that the sifted key is created only from signals within the same block. To guarantee this, one could discard those detection events where Bob declares time instances that belong to different blocks. Alternatively, one could change Bob's public announcement slightly. For example, one can reorder the $2m$ possible detection time slots of a given block to form a closed chain with the first and last time instances connected. Now, if Bob observes a ``click" in the data line in say the  first time slot he announces a detection event in time instances one and three or one and $2m-1$ with equal probability, and similar for the other cases. This strategy preserves the original symmetry in Bob's announcement and we use it in our simulations.

\begin{figure}[!t]\center
\vspace{-0.3cm}
\includegraphics[angle=-90,scale=0.68]{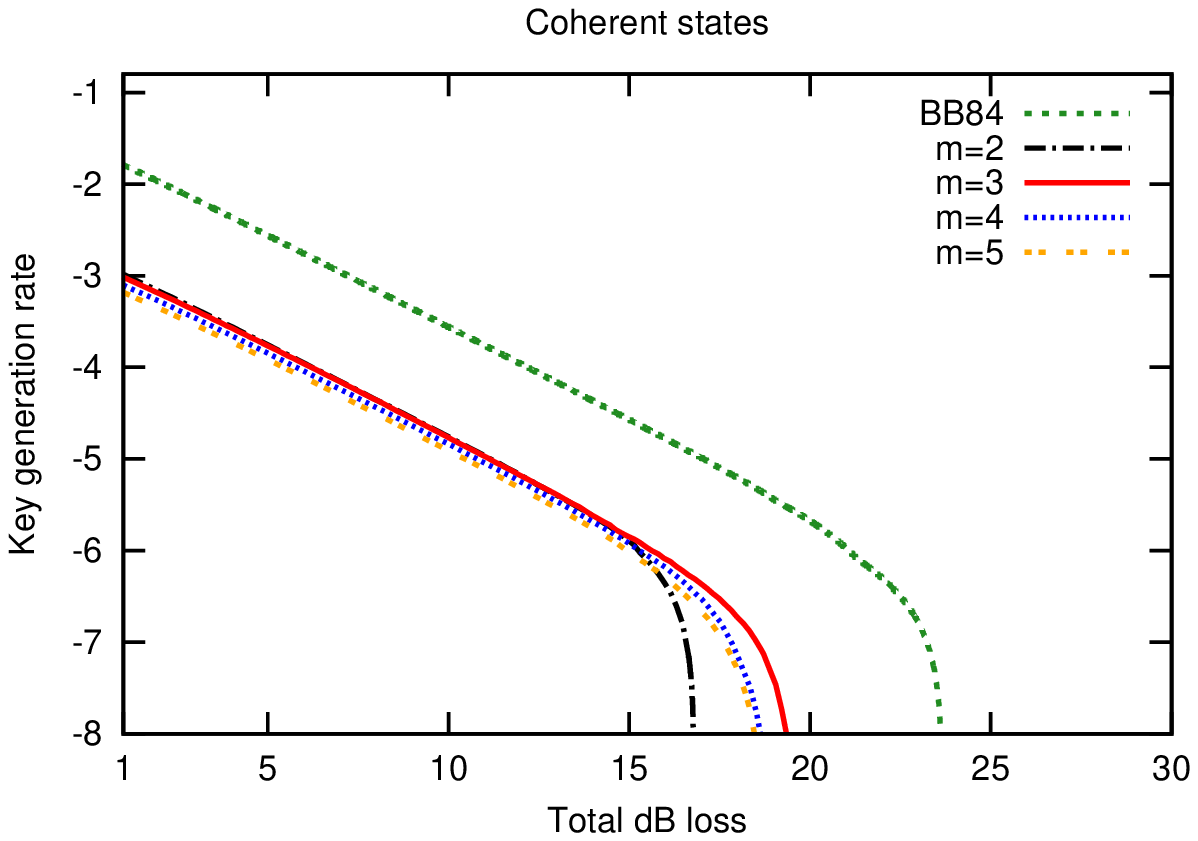}
\includegraphics[angle=-90,scale=0.68]{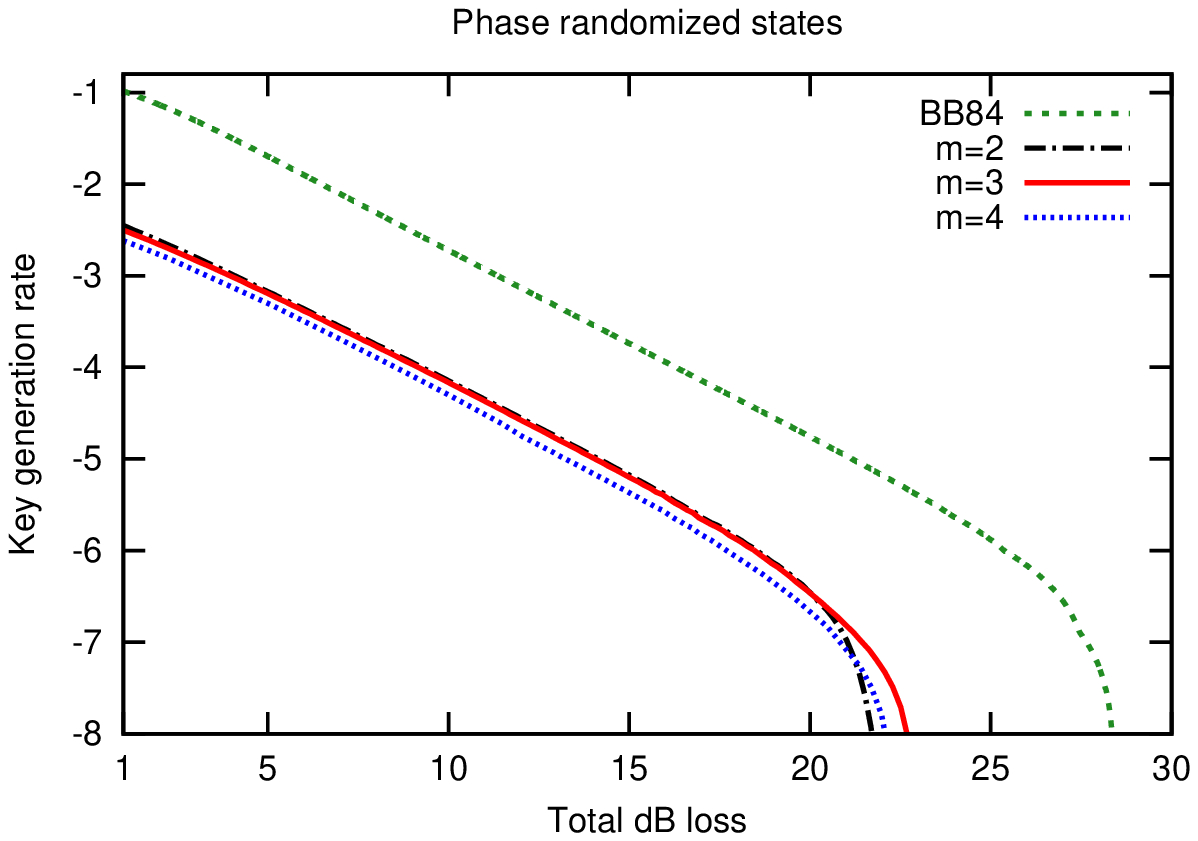}
\caption{Lower bound on the secret key rate given by Eq.~(\ref{eq:bound_average}) per pulse on a logarithmic scale (base 10) vs. the total system loss in dB for the COW protocol illustrated in Fig.~\ref{COWfig} using signal blocks carrying $m$ bits of information (\ie, $2m$ optical pulses) in the security proof. The upper figure corresponds to the case where all blocks of signals share a common phase, while the lower figure represents the situation where each block is phase randomized. For comparison, we include a lower bound on the secret key rate for a coherent-state version of the standard BB84 protocol \cite{BB84} with and without phase randomization \cite{gllp,lo_preskill}. We consider three main error contributions: an intrinsic error rate of $1\%$ in the data line, an additional misalignment in the monitoring line reducing the visibility to $99\%$, and a dark count rate in the detectors of $10^{-7}$. Moreover, in the lower figure we assume $n_{\rm cut}=2$. \label{fig_rates}}
\end{figure}

\textit{Simulation.}---
For simulation purposes, we consider that Bob's detectors are identical and have a dark count rate of $10^{-7}$. The channel model includes an intrinsic error rate of $1\%$ in the data line together with an additional misalignment in the monitoring line that reduces the interferometric visibility  to $99\%$. More details on this channel model and on the adapted security discussion to the COW protocol are given in the appendix. We study two different scenarios: (a) the case where all different $m$-signals blocks share the same phase, and (b) the scenario where each block is phase randomized. The resulting lower bounds on the secret key rate per pulse, \ie, $R_m/(2m)$, are illustrated in Fig.~\ref{fig_rates}. For comparison, this figure includes as well a lower bound on the secret key rate for a coherent-state version of the standard BB84 protocol~\cite{BB84} with and without phase randomization~\cite{gllp,lo_preskill}. For a given total system loss, \ie, including the losses in the channel and in Bob's detection apparatus, we optimize the lower bound over the respective signal strength $\alpha$ of Alice's source which is of order $0.1$. As expected, we find that case (b) performs better than that where all blocks share a common phase, since the signal states are less distinguishable for an eavesdropper without a global phase. We obtain that the tolerable system loss for the COW protocol is, respectively, $\approx19.5$ dB (a), and $\approx{}22.6$ dB (b). The bit error and visibility at these cutoff points are, respectively, $\approx3\%$ and $\approx96\%$ (a), and $\approx5.3\%$ and $\approx93.3\%$ (b). Let us remark that the lower bound with $m=2$ even holds for threshold detectors~\cite{note}.

Our simulations reveal that a main limiting factor in DPR-QKD seems to be the dark count rate of Bob's detectors. For given experimental parameters, there is an optimal finite block size that allows a maximum tolerable total system loss. If one increases the block size further this does not translate into an improved lower bound or distance. This is due to the fact that, in the high loss regime, large sized blocks suffer from a higher dark count probability {\it per block} than smaller sized blocks, and this reduces the achievable secret key rate. A similar effect was already observed in the security analysis for the differential-phase-shift protocol with true single photon sources~\cite{kiyo}. For a dark count rate per pulse of $10^{-7}$ the optimal block size in the COW scheme turns out to be $m=3$, \ie, $6$ optical pulses. Also, this figure shows that a coherent-state version of the BB84 protocol without decoy states can deliver notably higher key rates per signal than the analyzed COW protocol assuming the same channel model. The reason for this might be threefold: (1) the small optimal block size in the COW scheme; (2) considering blocks, it can be shown that certain multi-photon pulses are completely insecure; (3) most importantly, while in the BB84 the phase error is measured directly, in the COW protocol it has to be estimated.

\textit{Possible improvements.}---
To further improve the lower bounds shown in Fig.~\ref{fig_rates} there are several alternatives. Since a main limitation seems to come from dark counts, one may consider security in the fully calibrated device scenario where these errors are not attributed to Eve. As a quantitative bound on the performance of this scenario we investigated the case of a zero dark count rate, in which all key rate bounds shown in Fig.~\ref{fig_rates} shift by about $3$ dB, though the difference between the COW and the BB84 protocol remains. Additionally, one can evaluate different public announcements in a similar spirit like the SARG protocol~\cite{sarg}. We considered different declarations, but unfortunately none of them enhanced the resulting key rate~\cite{announcements}. Another possibility is to include, for instance, an extra monitoring line on Bob's side to additionally check the coherence between subsequent pulses. The state distribution part of this protocol is then very similar to the one of the original COW scheme~\cite{cow} with an additional decoy signal composed by two vacuum pulses as proposed in Ref.~\cite{upper_bound_cow}. This hardware change improves the maximum tolerable system loss by about $1$~dB. 

Another hardware change might be to include additional phase differences in the signal stream, such that the signals states get closer to the one used in a BB84 protocol. Finally, one may ask whether different security techniques might provide better lower bounds. For instance, one could consider more valid detection events per block. This needs however much larger block sizes such that one obtains at all a reasonable fraction of two or more click events in the long distance limit. Another alternative would be to bound the rate by the individual phase errors, \ie, directly using Eq.~(\ref{eq:bound_individual}). This could give a benefit if, for example, bits at the boundary are much easier to infer by Eve than bit values originating from events well inside the block. Moreover, it might be of advantage if Eve's information is estimated by using different, possibly not mutually unbiased basis measurements. Here the more general key rate formula of Ref.~\cite{lower_bounds} could be used. Clearly another option would be to abandon the block idea. However even in this case Eve could always attack the signals block-wise. Though a coherence measurement across blocks would then reveal the eavesdropper, any coherence measurement within them would be still fine. Hence when considering only an average visibility this effect will become less and less important. All these alternatives definitely deserve further investigations, but we do not expect a dramatic improvement.

\textit{Conclusion.}---
We have presented a generic method to prove security of practical DPR-QKD against general attacks. With the explicit example of a variant of the COW protocol, we have shown that these schemes are indeed secure for certain distances at given rates. Its performance, however, seems to be less robust against practical imperfections than originally expected.  

\begin{acknowledgments}
We would like to thank H.-K. Lo, N.~L\"utkenhaus, V.~Scarani, L.~Sheridan and N.~Walenta for stimulating discussions about the topic and technicalities, and L.~M. Eriksson for comments on the presentation of the paper. T.M., M.C., and L.P.T. especially thank the Group of Applied Physics, University of Geneva, for hospitality and support during their stay at this institution, where parts of this research have been conducted. This work has been supported by the EU (Marie Curie CIG 293993/ENFOQI), the BMBF (Chist-Era Project QUASAR), the National Research Foundation and the Ministry of Education, Singapore, the National Centre of Competence in Research QSIT, the Swiss NanoTera project QCRYPT and the FP7 Marie-Curie IAAP QCERT project.  
\end{acknowledgments}

\bibliographystyle{apsrev}

\begin{thebibliography}{99}
\bibitem{note_friday} Sequential attacks~\cite{dps_sequential,upper_bound_cow} against DPR-QKD also show that its key generation rate differs from the one obtained if one assumes only collective attacks.
\bibitem{shield} A shield system refers to an ancillary system that is inaccessible to the eavesdropper. Moreover let us point out that though a single purification $\ket{\rho_i^m}_{\rm A_sB}$ is unique up to local unitary, here one requires that all signals $\rho_i^m$ are purified to the same shield system $A_{\rm s}$, which is not unique anymore. While certain collective purifications are clearly better than others, any choice is valid.
\bibitem{comment_active} In our simulation we consider, for simplicity, an active measurement choice by Bob. In practice, this can be replaced by a passive optical coupler to select between data and monitoring line, together with an unbalanced Mach-Zehnder interferometer for the coherence measurement.
\bibitem{note}
In the easiest setting Bob may want to use threshold detectors, \ie, simple click/no-click detectors, instead of photon number resolving detectors. Using similar arguments like in Ref.~\cite{squash}, it can be shown that, for 2-bit blocks, Bob's measurement apparatus can be described with a squash model and the security analysis applies directly. The resulting secret key rate, however, is almost identical to the case where he uses photon number resolving detectors, only slightly higher in the low loss regime where the probability to obtain a valid detection event is also higher.
\bibitem{announcements} In particular, we have examined two further cases. In the first one, whenever 
Bob sees a ``click" in detector ${\rm D}_d$ in time instance $i\in\{1,2\}$ ($i\in\{2m,2m-1\}$) he always declares a detection 
event in time instances $i$ and $i+2$ ($i$ and $i-2$). Detection events not situated 
in the border of the blocks are treated as described in the main text. In the second strategy, 
detection events produced in the border of the blocks are only announced by Bob with probability $1/2$. Both 
methods deliver lower key rates 
than the one described in the main text.
\end{thebibliography}

\appendix

\section*{Appendix}

In this appendix we apply the described security method to the explained version of the COW protocol~\cite{private}. In particular, we consider signal blocks carrying $m$~bits of information. Since a single bit comprises two modes, one has $2m$ different temporal modes described by their creation and annihilation operators $a_s^\dag$ and $a_s$, respectively, with $s=1,\dots, 2m$. We assume that the $l$-th bit relates to the modes with $s=2l-1,2l$.

\begin{figure}[!h]\center
\resizebox{6.2cm}{!}{\includegraphics{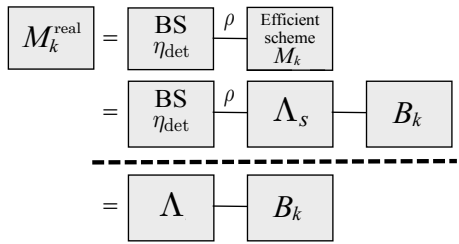}} 
\caption{Decomposition of Bob's measuring device.}
\label{fig:measurements}
\end{figure}

\textit{Real and assumed measurement description.}---At first let us concentrate on the real measurement model $M_k^{\rm real}$ and the way how we describe it in the security part, denoted as $B_k$ in the main text. For the real measurement setup we assume inefficient photon number resolving detectors that suffer from state-independent dark counts. The inefficiency of $M_k^{\rm real}$ is modelled by a global beamsplitter (BS) of transmittance $\eta_{\rm det}$ located in front of a perfectly efficient scheme, labelled as $M_k$, that still suffers from dark counts. This is schematically drawn in the first line of Fig.~\ref{fig:measurements}. In a second step, one models the efficient scheme $M_k$ as a map $\Lambda_{\rm s}$, sometimes called squashing or filter operation~\cite{squash}, in front of the assumed description $B_k$. Let us emphasize that the security simulation is valid for any true measurement scheme that can be modelled as a physical map $\Lambda$ followed by the measurement $B_k$ as shown in the third line of the figure.

There are three different types of outcomes for the so far abstract outcome label ``$k$''. For a data line measurement we use $d$, with $d=1,\dots, 2m$, to denote a single photon detection in temporal mode $d$ only. The corresponding measurement operator $M_d$ is given by 
\begin{equation}
M_d = \epsilon (1-\epsilon)^{2m-1} \ket{\rm vac}\bra{\rm vac} + (1-\epsilon)^{2m} \ket{d}\bra{d},
\end{equation}
with $\epsilon$ representing the dark count probability of Bob's detectors and $\ket{d}=a_d^\dag \ket{\rm vac}$. In addition to a data line measurement Bob can also perform coherence measurements on subsequent bits employing the monitoring line. For instance, whenever he tests the coherence between bits $l$ and $l+1$ he effectively mixes the modes $2l-1,2l+1$ and, at the same time, $2l,2l+2$. For each pair of modes there are two single photon events, denoted as $\pm$, that can be distinguished, depending on whether the single excitation is registered in the bright (${\rm D}_+$) or in the dark (${\rm D}_-$) detector. As an outcome label for the coherence measurements we use $k=(c,\pm)$, where $c=1,\dots,2m-2$ denotes the first of the two interfering modes. In this case the measurement operators are given by
\begin{equation}
M_{c,\pm} = \epsilon (1-\epsilon)^{2m-1} \ket{\rm vac}\bra{\rm vac} + (1-\epsilon)^{2m} \ket{\chi_c^{\pm}}\bra{\chi_c^\pm},
\end{equation}
with $\ket{\chi_c^{\pm}}=(\ket{c}\pm\ket{c+2})/\sqrt{2}$. Let us emphasize that in these coherence measurements it is still necessary to check that all other modes are empty. Finally, note that each measurement setting has also other possible outcomes, \eg, ``no click" or more than a single photon detection event. All these cases are grouped (via classical post-processing) into a single inconclusive outcome described by $M_{\rm inc}$.

As the modelled measurement operators $B_k$ we use
\begin{eqnarray}
B_d &=& \ket{d}\bra{d}, \nonumber \\
B_{c,\pm} &=&  \ket{\chi_c^{\pm}}\bra{\chi_c^\pm}, \nonumber\\
B_{\rm inc} &=& \ket{a}\bra{a},
\end{eqnarray}
where $\ket{a}$ is the auxiliary state that describes the inconclusive outcome. These measurement operators $B_k$ act on a $2m+1$ dimensional Hilbert space. 

Both measurement sets can be made equivalent by an appropriate map $\Lambda_s$ such that $\tr(\rho M_k)=\tr[\Lambda_s(\rho) B_k]$ holds for all possible states $\rho$ and measurement outcomes ``$k$'' as schematically shown in Fig.~\ref{fig:measurements}. This map $\Lambda_s$ is given as follows. First one measures the total number of photons $n$ within an arriving block. Whenever one finds $n\geq 2$ one outputs the auxiliary state $\ket{a}$. If $n=1$ then with probability $(1-\epsilon)^{2m}$ the single photon state stays untouched, otherwise the auxiliary state is thrown again. Finally, for $n=0$ the map creates the completely mixed single photon state $\sum_k \ket{k}\bra{k}/2m$ with probability $2m \epsilon (1-\epsilon)^{2m-1}$ and $\ket{a}$ otherwise. This map is physical because we explicitly describe it in terms of measurements and conditional signal state preparations.

\textit{Source state and reduced density matrix.}---
The following discussion provides the source states for both cases of pure or phase randomized COW block signals. These states determine the reduced density matrix $\rho_{\rm A}^m$ which belongs to the available information. 

Let us consider first the case of pure signal states. In the COW protocol analyzed Alice sends to Bob either the sequence $\ket{\alpha,0}$ or $\ket{0,\alpha}$, with $\alpha \in \mathbbm{R}$, depending on whether her raw key bit value is ``$0$" or ``$1$". Let us start with the scenario where Alice sends to Bob only one bit value, occurring with equal a priori probability. This corresponds to a block size $m=1$. Then the source state is given by 
\begin{equation}
\ket{\Psi^{m=1}}_{\rm AB} =\frac{1}{\sqrt{2}} \big( \ket{0}_{\rm A} \ket{\alpha,0}_{\rm B} + \ket{1}_{\rm A} \ket{0,\alpha}_{\rm B}  \big),
\end{equation} 
and its reduced density matrix becomes
\begin{equation}
\rho_{\rm A}^{m=1} =\frac{1}{2} \left[ \begin{array}{cc} 1 & e^{-\alpha^2} \\ e^{-\alpha^2}  & 1 \end{array} \right].
\end{equation} 

Suppose now that Alice sends to Bob $m$ bits according to this scheme. If $i=(i_1,i_2,\dots,i_m)$ denotes the $m$-bit string being sent and $\ket{\phi_i}_{\rm B}$ refers to the corresponding signal state, then one obtains
\begin{eqnarray}
\ket{\Psi^{m}}_{\rm AB} &=& 2^{-\frac{m}{2}} \sum_{i \in \{ 0,1\}^m} \ket{i}_{\rm A} \ket{\phi_i}_{\rm B} \nonumber \\
&=&\ket{\Psi^{m}}_{\rm A_1\dots A_mB} = \ket{\Psi^{m=1}}_{\rm AB}^{\otimes m}.
\end{eqnarray}
In particular, from the last expression one finds that the reduced density matrix $\rho_{\rm A}^m$ is given by
\begin{equation}
\rho_{\rm A}^m = (\rho_{\rm A}^{m=1})^{\otimes m}.
\end{equation}

Next, let us turn to the case of phase randomized blocks. Since randomizing the phase of a block is equivalent to measuring the total number of photons contained in it, the true signals states are of the form 
\begin{equation}
\rho_i^m = \sum_{n=0}^\infty \Pi_n \ket{\phi_i}_{\rm B}\bra{\phi_i} \Pi_n = \sum_{n=0}^\infty p_\lambda (n) \ket{\psi^i_n}_{\rm B}\bra{\psi^i_n}.
\end{equation}
Here $\Pi_n$ stands for the projector onto the $n$-photon subspace of the $2m$ different modes. The outcome of such a photon number measurement follows a Poisson distribution $p_\lambda (n) = e^{-\lambda}\lambda^n/n!$ with mean $\lambda=m \alpha^2$. The projected $n$-photon signal states $\ket{\psi^i_n}_{\rm B}$ can be expressed as
\begin{equation}
\ket{\psi^i_n}_{\rm B} = m^{-\frac{n}{2}} \sqrt{n!}  \sum_{n_1,\dots n_m} \prod_{l=1}^m \frac{(a_{2l+i_l-1}^\dag)^{n_l}}{n_l!} \ket{\rm vac}_{\rm B},
\end{equation}
where the summation runs over all natural numbers $n_1,\dots,n_m$ that satisfy $\sum_{l=1}^m n_l = n$. These states fulfill the relation 
\begin{equation}
\label{eq:identity}
\braket{\psi^i_n}{\psi^j_{\bar n}} = \delta_{n \bar n} \left( \frac{m- \Delta_{ij}}{m}\right)^n,
\end{equation}
with $\Delta_{ij}$ being the Hamming distance between the bit strings $i$ and $j$, \ie, the number of places they differ.

Using the framework of mixed signal states as explained in the main text one must now choose an overall purification of all signal states 
$\ket{\psi^i_n}_{\rm B}$. For our simulation we select
\begin{equation}
\label{eq:purifications}
\ket{\rho_i^m}_{\rm A_sB} = \sum_{n=0}^\infty \sqrt{p_\lambda(n)} \ket{n}_{\rm A_s} \ket{\psi_n^i}_{\rm B},
\end{equation}
which can be seen as a coherent storage of the total photon number $n$ in the shield system $\rm A_s$. Let us remark that this choice satisfies  $\braket{\rho_i^m}{\rho_j^m}=F(\rho_i^m,\rho_j^m)$, with $F$ being the fidelity of mixed states, which is also the maximal possible overlap between two signal states~\cite{fidelity}. We find, therefore, that the source state in this scenario is given by
\begin{equation}
\ket{\Psi^m}_{\rm A_bA_sB} = 2^{-\frac{m}{2}} \sum_{i \in \{ 0,1\}^m} \ket{i}_{\rm A_b} \ket{\rho_i^m}_{\rm A_sB},
\end{equation}
with $\rm A_b={\rm A_1\dots A_m}$.
This means that the reduced density matrix $\rho_{\rm A}^m$, with $\rm A=A_bA_s$, can be expressed as
\begin{eqnarray}
\rho_{\rm A}^m &=& \sum_{n=0}^\infty p_\lambda(n) \rho_{\rm A_b}^n \otimes \ket{n}_{\rm A_s}\bra{n}, 
\end{eqnarray}
with $\rho_{\rm A_b}^n$ given by
\begin{equation}
\rho_{\rm A_b}^n = 2^{-m} \sum_{i,j} \left( \frac{m- \Delta_{ij}}{m}\right)^n \ket{i}_{\rm A_b}\bra{j}.
\end{equation}

In our simulation we only use partial information of the reduced density matrix $\rho_{\rm A}^m$. In particular, we transform $\rm A_s$ to $\rm \bar{A}_s$ by making a shield measurement that distinguishes the different photon number cases mentioned in the main text such that one obtains
\begin{eqnarray}
\label{eq:tagged_rhoA}
\rho_{\rm A_b \bar {\rm A_s}}^m &=& \sum_{n=1}^{n_{\rm cut}} p_\lambda(n) \rho_{\rm A_b}^n \otimes \ket{n}_{\rm \bar{A}_s}\bra{n} \nonumber\\
&& \;\; + \sum_{n \not \in \{1,\dots,n_{\rm cut}\}} p_\lambda(n) \rho_{\rm A_b}^n \otimes \ket{N}_{\rm \bar{A}_s}\bra{N},
\end{eqnarray}
where $\ket{N}_{\rm \bar{A}_s}$ denotes an auxiliary system for all higher photon numbers. Let us point out that considering the reduced state given by Eq.~(\ref{eq:tagged_rhoA}) can be understood as ``tagging'' the $n=1,\dots,n_{\rm cut}$ signal states~\cite{gllp}.

\textit{Announcement maps and phase operator.}---The specific announcements $v$ of the COW protocol can be phrased in terms of appropriate maps $\Lambda_v$ on the quantum state. Together with a chosen ``phase setting'' measurement this provides a concrete expression for the averaged phase error operator $F_{\bar \delta}$ used in Eq.~(\ref{eq:phase_error_operator}). 

As explained in the protocol description, Bob announces two consecutive even or odd time slots where he registered his single photon event. Suppose, for instance, that he announces $v=(2l-1,2l+1)$. These are the first arrival times of the modes associated with bits $i_l$ and $i_{l+1}$ sent by Alice. In such cases, Alice and Bob agree to call the outcome in the first time instance~``$0$'' while the later event is~``$1$''. This announcement can be modelled as a filter operation $\Lambda^{\rm B}_v(\rho)=F^{\rm B}_v \rho F^{{\rm B} \dag}_v$ given by
\begin{equation}\label{f1s}
F^{\rm B}_v =\frac{1}{\sqrt{2}} \left( \ket{0}_{\rm \bar B B} \bra{2l-1} + \ket{1}_{\rm \bar B B} \bra{2l+1} \right).
\end{equation}
If Bob measures system $\bar {\rm B}$ in the standard basis $\ket{0}_{\rm \bar B},\ket{1}_{\rm \bar B}$ he obtains the real outcome he has observed. The pre-factor $1/\sqrt{2}$ which appears in Eq.~(\ref{f1s}) takes into account that whenever Bob sees a single photon click in either $2l-1$ or $2l+1$ he announces this particular $v$ with just $50\%$ probability, \ie, $F^{{\rm B} \dag}_v F^{{\rm B}}_v = (B_{2l-1}+B_{2l+1})/2$.

\begin{figure}[!h]\center
\resizebox{6.5cm}{!}{\includegraphics{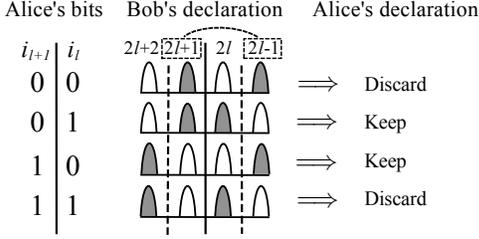}} 
\caption{Announcement choices for Alice given that Bob has declared a detection event in time slots $2l-1$ and $2l+1$.}
\label{fig:announcements}
\end{figure}

Suppose Bob has actually declared $v=(2l-1,2l+1)$. Then, Alice has to look on her bit string to determine whether she can conclusively infer Bob's bit value. For that, only her bits $i_l$ and $i_{l+1}$ matter. As shown in Fig.~\ref{fig:announcements}, if these two bits are equal it means that she had sent to Bob either two full or two empty pulses. In this scenario, she cannot infer Bob's bit value and they discard this result. However, if these bits differ then she knows Bob's sifted bit value precisely (in the error free case) and she tells Bob to keep it. 
Such a conclusive announcement by Alice can similarly be modelled as a filter operation $\Lambda^{\rm A}_v$ acting on her qubits $l$ and $l+1$, \ie, $\Lambda^{\rm A}_v(\rho^m_{\rm A_1\dots A_m}) =F^{\rm A}_v \rho^m_{\rm A_l A_{l+1}} F^{{\rm A} \dag}_v$ with
\begin{equation}
F^{\rm A}_v = \ket{0}_{\rm \bar A\:\: A_l A_{l+1}}\!\bra{01} + \ket{1}_{\rm \bar A\:\: A_l A_{l+1}}\!\bra{10}.
\end{equation}
Again a measurement in the standard basis $\ket{0}_{\rm \bar A},\ket{1}_{\rm \bar A}$ provides Alice with her real outcomes.

In order to determine the phase error $\delta_{v}$ we assume that both parties perform measurements in the $X$-basis, \ie, they project the output signals from their filter operations onto the states $\ket{\pm}=(\ket{0}\pm\ket{1})/\sqrt{2}$. Then, the symmetrized phase error $\delta_{v}=p(+,-)+p(-,+)$ can be expressed as
\begin{eqnarray}\label{spit}
p(v) \delta_v &=& p(v) \tr [ \frac{1}{2}  (\mathbbm{1} \otimes \mathbbm{1}-\sigma_x \otimes \sigma_x ) \sigma^{m}_{{\rm \bar A \bar B},v} ] \nonumber\\
&=&\frac{1}{2} p(v) - \frac{1}{2} \tr( \sigma_x \otimes \sigma_x p(v) \sigma_{{\rm \bar A \bar B},v}^{m} ) \nonumber\\
&=& \frac{1}{2} p(v) - \tr( X^\prime_{\rm A}\otimes X^\prime_{\rm B}  \rho^m_{\rm AB}),
\end{eqnarray}
with $\sigma_x$ denoting the Pauli matrix $\sigma_x=\ket{0}\bra{1}+\ket{1}\bra{0}$.
In the last line of Eq.~(\ref{spit}) we have defined the operators
\begin{eqnarray}
X^\prime_{\rm A} &=& \mathbbm{1}_{\rm A_1\dots A_{l-1}} \otimes X_{\rm A} \otimes \mathbbm{1}_{\rm A_{l+2}\dots A_m},\nonumber\\
\nonumber 
X^\prime_{\rm B} &=& \frac{1}{2} F_v^{{\rm B} \dag} \sigma_x F_v^{\rm B}\\ &=& \frac{1}{4} \left( \ket{2l-1}\bra{2l+1} + \ket{2l+1}\bra{2l-1} \right),
\end{eqnarray}
with $X_{\rm A} = F_v^{{\rm A} \dag} \sigma_x F_v^{\rm A} =\ket{01}\bra{10}+\ket{10}\bra{01}$. 

Similar arguments apply to the cases where Bob announces subsequent even outcome pairs or the special instances at the borders of the blocks. We find that the averaged phase error $\bar \delta =\sum_{v \in \mathcal{V}_{\rm c}} p(v) \delta_v$ can be written as
\begin{equation}
\bar \delta = \frac{1}{2} \sum_{v \in \mathcal{V}_{\rm c}} p(v) - \tr( X_{\bar \delta} \rho_{\rm AB}^m),
\end{equation}
with an operator $X_{\bar \delta} = \sum_{l=1}^m X_{{\rm A};l} \otimes X_{{\rm B};l}$. Here $X_{{\rm A};l}$ denotes the operator composed by the previously defined $X_{\rm A}$ acting on qubits $l$ and $l+1$ and the identity operator acting on the remaining qubits ($l=m$ means the first and last qubit). On Bob's side the operators $X_{{\rm B};l}$ are given by
\begin{eqnarray}
\nonumber
X_{{\rm B};l}=\frac{1}{4} (&& \ket{2l-1}\bra{2l+1} + \ket{2l+1}\bra{2l-1} \\ 
&& + \ket{2l}\bra{2l+2} + \ket{2l+2}\bra{2l}),
\end{eqnarray}
with addition being carried out modulo $2m$.

\textit{Channel model.}---In this section we present the employed channel model of the COW experiment used in our numerical simulations. Note, however, that the results presented in this article can be applied as well to any other quantum channel, as they only depend on the observed detection probabilities in both the data and monitoring lines. 

In particular, we characterize the losses in the channel with a BS of transmittance $\eta_{\rm channel}$. This parameter can be related with a transmission distance $d$ measured in km for the given QKD scheme as 
\begin{equation}
\eta_{\rm channel}=10^{-\frac{\alpha{}d}{10}},
\end{equation}
where $\alpha$ represents the loss coefficient of the channel (\eg, an optical fiber) measured in dB/km. Together with the efficiency of the detectors the overall system transmittance is given by
\begin{equation}
\eta_{\rm sys}=\eta_{\rm channel}\eta_{\rm det}.
\end{equation}
The total system loss in dB is used as the x-axis in the secret key rate figures, \ie, $-10\log_{10}\eta_{\rm sys}$.

The channel misalignment is parametrized with an error probability $e_{\rm d}$ that a signal hits Bob's detectors in the wrong time slot within the same bit. For simplicity, we assume that $e_{\rm d}$ is a constant independent of the distance and we use $e_{\rm d}=1\%$ for simulation purposes. This effect is modelled by a completely positive trace-preserving map $\Phi$ that incoherently flips the signal states within the same bit slot as $\ket{0,\sqrt{\eta_{\rm sys}}\alpha}\mapsto\ket{\sqrt{\eta_{\rm sys}}\alpha,0}$ and $\ket{\sqrt{\eta_{\rm sys}}\alpha,0}\mapsto\ket{0,\sqrt{\eta_{\rm sys}}\alpha}$ with probability $e_{\rm d}$. Here we consider that the input signals have been already affected by  system losses. We have, therefore, that whenever Alice sends to Bob a corresponding COW signal state with coherent state $\ket{\alpha}$ in temporal mode $d$, the probability that Bob observes a single photon detection event in this mode only (within the whole signal block) is given by
\begin{eqnarray}
p_{d}^{\rm correct}&=&\tr\{\Lambda_s[\Phi^{\otimes{}m}({\rho}^m_{\rm loss})]M_d\}=\epsilon(1-\epsilon)^{2m-1}e^{-\eta_{\rm sys}\lambda} \nonumber \\ 
&&+(1-\epsilon)^{2m}(1-e_{\rm d})\eta_{\rm sys}\mu{}e^{-\eta_{\rm sys}\lambda},
\end{eqnarray}
where $\rho_{\rm loss}^m$ represents the output signal of the BS characterizing the total system loss, $\mu=\alpha^2$, and $\lambda=m \mu$. Similarly, when Alice sends to Bob a vacuum state in temporal mode $d$ Bob can observe a single photon detection event in this mode only with probability
\begin{eqnarray}
p_{d}^{\rm error}&=&\epsilon(1-\epsilon)^{2m-1}e^{-\eta_{\rm sys}\lambda} \nonumber \\ 
&&+(1-\epsilon)^{2m}e_{\rm d}\eta_{\rm sys}\mu{}e^{-\eta_{\rm sys}\lambda}.
\end{eqnarray}
The total probability that Bob observes an inconclusive detection event in the data line is then given by
\begin{eqnarray}
p_{\rm inc}=1-m\big(p_{d}^{\rm correct}+p_{d}^{\rm error}\big).
\end{eqnarray}

In the monitoring line we include an additional misalignment effect that reduces further the interferometric visibility. In particular, we assume that whenever two equal coherent states interfere at a $50:50$ BS then the outcome signal can exit the BS through the wrong output port with error probability $e_{\rm m}$. In our simulations we use $e_{\rm m}=0.5\%$. Here we distinguish two possible scenarios, depending on whether the signals which interfere at the BS were prepared by Alice in the same quantum state or not. Let us assume that the first signal corresponds to bit $i_l$ while the later to bit $i_{l+1}$. That is, Bob interferes modes $2l-1,2l+1$ and, at the same time, $2l,2l+2$.

Let us consider first the situation where both signals were generated in the same state $\ket{0,\alpha}$. In this scenario, we find that Bob observes a single photon detection event in temporal mode $2l-1$ only (and no click in the remaining modes of the block) with probability
\begin{eqnarray}\label{trd1}
p_{2l-1,+}&=&\epsilon(1-\epsilon)^{2m-1}e^{-\eta_{\rm sys}\lambda}+(1-\epsilon)^{2m}\eta_{\rm sys}\mu{}e^{-\eta_{\rm sys}\lambda} \nonumber \\
&&\times\left[2(1-e_{\rm d})^2(1-e_{\rm m})+e_{\rm d}(1-e_{\rm d})\right], \nonumber \\
p_{2l-1,-}&=&\epsilon(1-\epsilon)^{2m-1}e^{-\eta_{\rm sys}\lambda}+(1-\epsilon)^{2m}\eta_{\rm sys}\mu{}e^{-\eta_{\rm sys}\lambda} \nonumber \\
&&\times\left[2(1-e_{\rm d})^2e_{\rm m}+e_{\rm d}(1-e_{\rm d})\right],
\end{eqnarray}
where the superscript $\pm$ indicates whether the single excitation is registered in the bright (${\rm D}_+$) or in the dark (${\rm D}_-$) detector of the monitoring line. Similarly, we have that the probability that Bob sees a single photon detection in temporal mode $2l$ only is given by
\begin{eqnarray}\label{trd2}
p_{2l,+}&=&\epsilon(1-\epsilon)^{2m-1}e^{-\eta_{\rm sys}\lambda}+(1-\epsilon)^{2m}\eta_{\rm sys}\mu{}e^{-\eta_{\rm sys}\lambda} \nonumber \\
&&\times\left[2e_{\rm d}^2(1-e_{\rm m})+e_{\rm d}(1-e_{\rm d})\right], \nonumber \\
p_{2l,-}&=&\epsilon(1-\epsilon)^{2m-1}e^{-\eta_{\rm sys}\lambda}+(1-\epsilon)^{2m}\eta_{\rm sys}\mu{}e^{-\eta_{\rm sys}\lambda} \nonumber \\
&&\times\left[2e_{\rm d}^2e_{\rm m}+e_{\rm d}(1-e_{\rm d})\right].
\end{eqnarray}
The case where both signals were generated in the same state $\ket{\alpha,0}$ is completely analogous. One only needs to interchange Eqs.~(\ref{trd1}) and (\ref{trd2}). 

Finally, let us consider the situation where both signals are prepared in a different quantum state. In this scenario the probabilities are given by
\begin{eqnarray}
\nonumber
p_{2l-1,+}&=&p_{2l,+} \\
\nonumber
&=& \epsilon(1-\epsilon)^{2m-1}e^{-\eta_{\rm sys}\lambda} +(1-\epsilon)^{2m}\eta_{\rm sys}\mu{}e^{-\eta_{\rm sys}\lambda} \\
&& \times \! \bigg[2e_{\rm d}(1-e_{\rm d})(1-e_{\rm m}) +\frac{1+2e_{\rm d}^2-2e_{\rm d}}{2}\bigg]\!,
\end{eqnarray}
and
\begin{eqnarray}
\nonumber
p_{2l-1,-}&=&p_{2l,-} \\
\nonumber
&=& \epsilon(1-\epsilon)^{2m-1}e^{-\eta_{\rm sys}\lambda} +(1-\epsilon)^{2m}\eta_{\rm sys}\mu{}e^{-\eta_{\rm sys}\lambda} \\
&& \times \! \bigg[2e_{\rm d}(1-e_{\rm d})e_{\rm m} +\frac{1+2e_{\rm d}^2-2e_{\rm d}}{2} \bigg]\!.
\end{eqnarray}

\end{document}